\documentclass{article}
\usepackage{graphicx}
\usepackage{amsmath}

\newtheorem{theorem}{Theorem}

\newtheorem{definition}[theorem]{Definition}

\input{tcilatex}
\begin{document}

\title{Power-Law Distributions:\\
Beyond Paretian Fractality}
\author{Iddo Eliazar\thanks{%
Department of Technology Management, Holon Institute of Technology, P.O. Box
305, Holon 58102, Israel. E-mail: \emph{eliazar@post.tau.ac.il}} \and Joseph
Klafter\thanks{%
School of Chemistry, Sackler Faculty of Exact Sciences, Tel Aviv University,
Tel Aviv 69978, Israel. E-mail: \emph{klafter@post.tau.ac.il}}}
\maketitle

\begin{abstract}
The notion of fractality, in the context of positive-valued probability
distributions, is conventionally associated with the class of Paretian
probability laws. In this research we show that the Paretian class is merely
one out of six classes of probability laws -- all equally entitled to be
ordained fractal, all possessing a characteristic power-law structure, and
all being the unique fixed points of renormalizations acting on the space of
positive-valued probability distributions. These six fractal classes are
further shown to be one-dimensional functional projections of underlying
fractal Poisson processes governed by: \textbf{(i)} a common elemental
power-law structure; and, \textbf{(ii)} an intrinsic scale which can be
either linear, harmonic, log-linear, or log-harmonic. This research provides
a panoramic and comprehensive view of fractal distributions, backed by a
unified theory of their underlying Poissonian fractals.

\bigskip


\bigskip

\end{abstract}

\section{\label{1}Introduction}

Fractal objects are ubiquitous across many fields of science, and their
study has attracted major interest by a broad array of researchers -- see 
\cite{Man}-\cite{Bar} and references therein. The geometric characteristic
of fractals is \emph{invariance under changes of scale}. The algebraic
manifestation of scale-invariance is given by \emph{power-laws}.

Power-laws facilitate the characterization of fractality when no geometry is
present -- the quintessential example being power-law probability
distributions (see Chapter \emph{38} in \cite{Man}).

Consider a population represented by a collection of points scattered
arbitrarily on the positive half-line -- the points representing the values
of the population members. Examples include: earthquakes taking place in a
given geological region, during a given period of time, measured by their
magnitudes -- each point representing the magnitude of an earthquake; stars
in a given sector of space measured by their masses -- each point
representing the mass of a star; citizens of a given state measured by their
wealth -- each point representing the wealth of a citizen; insurance claims
in a given insurance-portfolio measured by their costs -- each point
representing the cost of a claim; etc.

Such populations are discrete objects possessing no natural geometry -- and
hence no natural geometric characterization of fractality. The natural
setting for the analysis of such populations is statistical -- providing the
following conventional algebraic-statistic definition of fractality: \emph{a
population is fractal if its population-values and their
occurrence-frequencies are connected via a power-law}.

Shifting from the statistical perspective to the probabilistic perspective
one picks at random a member of the population, and considers its random
value $X$. Fractality, in the probabilistic setting, is characterized by a 
\emph{power-law survival probability }of the random variable $X$: 
\begin{equation}
\text{\textbf{Prob}}\left( \overset{\text{ }}{X>x}\right) =\left( \frac{a}{x}%
\right) ^{\alpha }  \label{1.01}
\end{equation}
($x>a$); the parameter $a$ being an arbitrary positive lower bound, and the
parameter $\alpha $ being an arbitrary positive exponent.

The probability distribution corresponding to the survival probability of
equation (\ref{1.01}) is referred to as \emph{Paretian} -- named after the
Italian economist Vilfredo Pareto who discovered, in 1896, a power-law
distribution of wealth in human societies \cite{Par}. The Paretian
probability distribution was empirically observed in a multitude of examples
coming from diverse scientific fields \cite{SZK}-\cite{KS} (see also the
review \cite{New} and references therein).

\bigskip

The theoretical construction of the Paretian power-law probability
distribution is based on the following pair of foundations: \textbf{(i)}
fractals are characterized, algebraically, by power-laws; \textbf{(ii)}
probability distributions are characterized, statistically, by survival
probabilities.

The first foundation implicitly assumes that \textquotedblleft
fractality\textquotedblright\ is synonymous with \textquotedblleft
power-laws\textquotedblright . This implicit assumption is \emph{false}. The
notion of fractality -- in the case of populations represented by
arbitrarily-scattered real-valued points -- can be defined from \emph{first
principles}. Namely, fractality can be defined via the elemental geometric
notion of \emph{scale-invariance} -- rather than via the emergent algebraic
notion of \emph{power-laws}. This approach, undertaken in \cite{EK1}, yields
three classes of \emph{non-Paretian} fractal populations.

The second foundation implicitly assumes that random variables are \emph{%
uniquely} characterized by their survival probabilities. This implicit
assumption is, again, false. Indeed, there are many ways of characterizing a
given probability distribution (these characteristics will be rigorously
defined in the sequel): Cumulative Distribution Functions (CDFs); Survival
Distribution Functions (SDFs); Backward Hazard Rates (BHRs); Forward Hazard
Rates (FHRs); Laplace Transforms (LTs); Moment Sequences (MSs); Log-Laplace
Transforms (LLTs); Cumulant Sequences (CSs).

Associating ``fractal distributions'' with power-law survival probabilities
yields Paretian probability distributions. But what if we associate
``fractal distributions'' with power-law Hazard Rates? or with power-law
Log-Laplace transforms? or with power-law Cumulants? This question serves as
the starting point of our research.

\bigskip

This paper is devoted to the exploration of the \emph{definition of
fractality} in the context of positive-valued probability distributions. As
we shall demonstrate, the notion of fractality is highly contingent on the
distribution-characteristic used. Defining fractality via power-law
structures of different distribution-characteristic leads to markedly
different probability distributions including: \emph{Pareto}, \emph{Beta}, 
\emph{Fr\'{e}chet}, \emph{Weibull}, \emph{L\'{e}vy Stable} -- all equally
entitled to be considered ``fractal distributions''.

Altogether we characterize six different classes of fractal distributions --
each class emerging from a power-law structure of a different
distribution-characteristic. Each of the six fractal classes characterized
is shown to be associated with a different \emph{renormalization}: the
members of each fractal class are the \emph{unique fixed points} of a
specific renormalization acting on the space of positive-valued probability
distributions. Each of the six fractal classes characterized is also
associated with a different \emph{Poissonian representation}: the members of
each fractal class are representable as a \emph{functional projection} of an
underlying class of Poisson processes defined on the positive half-line.

Having characterized the six different classes of fractal distributions,
their renormalizations, and their Poissonian representations, we turn to
seek an underlying \emph{unifying fractal structure}. To that end we study 
\emph{Poissonian renormalizations} -- renormalizations of Poisson processes
defined on the positive half-line -- and characterize four classes of \emph{%
Poissonian fractals}: the unique fixed points of multiplicative and
power-law Poissonian renormalizations.

The Poissonian fractals turn out to be governed by two structures: \textbf{%
(i)} a \emph{power-law} structure common to all Poissonian fractal classes; 
\textbf{(ii)} an \emph{intrinsic scale} which differentiates between the
four Poissonian fractal classes and characterizes them. The intrinsic scale
can be either linear, harmonic, log-linear, or log-harmonic. Moreover, the
Poissonian fractals turn out to be the Poisson processes underlying the
aforementioned fractal distributions. And, the ``algebraic fractality'' on
the ``probability-distribution level'' turns out to be a one-dimensional
projection of a more elemental ``geometric fractality'' prevalent on the
underlying ``Poisson-process level''.

This research provides a panoramic and comprehensive view of fractal
distributions, backed by a unified theory of their underlying Poissonian
fractals. The manuscript is organized as follows. The six classes of fractal
distributions -- as well as their associated renormalizations -- are
characterized in Section \ref{2}. Poissonian representations of the fractal
distributions are presented in Section \ref{3}. The Poissonian
renormalizations and Poissonian fractals underlying the six classes of
fractal distributions are unveiled in Section \ref{4}.

\bigskip

\textbf{Acronym glossary}

Throughout the manuscript the following acronyms shall be frequently used
(the subsections in brackets indicate the location, in the manuscript, of
the corresponding definitions):

\begin{description}
\item IID = Independent and Identically Distributed

\item PDF = Probability Density Function (Subsection \ref{2.1})

\item CDF = Cumulative Distribution Function (Subsection \ref{2.1})

\item SDF = Survival Distribution Function (Subsection \ref{2.1})

\item BHR = Backward Hazard Rate (Subsection \ref{2.2})

\item FHR = Forward Hazard Rate (Subsection \ref{2.2})

\item LT = Laplace Transform (Subsection \ref{2.3})

\item MS = Moment Sequence (Subsection \ref{2.3})

\item LLT = Log-Laplace Transform (Subsection \ref{2.3})

\item CS = Cumulant Sequence (Subsection \ref{2.3})

\item CRF = Cumulative Rate Function (Subsection \ref{3.0})

\item SRF = Survival Rate Function (Subsection \ref{3.0})
\end{description}

The equality sign $\overset{\text{Law}}{=}$ shall henceforth denote equality
in law (of random variables).

\section{\label{2}Power-law characterization of fractal distributions}

As noted in the introduction, positive-valued probability distributions have
various distribution-characteristics. In general, a
distribution-characteristic $\mathcal{C}_{D}$ of a positive-valued
probability distribution $D$ is a function $\mathcal{C}_{D}=\mathcal{C}%
_{D}(\theta )$ ($\theta \in \Theta $; $\Theta $ being a subset of the
non-negative half line) which uniquely determines $D$.

The aim of this research is to explore the notion of \emph{fractality}, in
the context of positive-valued probability distributions, via the following
definition:

\begin{definition}
A probability distribution $D$ is $\mathcal{C}$-fractal if its
distribution-characteristic $\mathcal{C}_{D}$ admits a power-law functional
structure: 
\begin{equation}
\mathcal{C}_{D}(\theta )=c\theta ^{\gamma }  \label{2.01}
\end{equation}
($\theta \in \Theta $), where $c$ is a positive coefficient and where $%
\gamma $ is a real exponent.
\end{definition}

Often, fractality is the manifestation of some underlying \emph{%
renormalization}. In the context of positive-valued probability
distributions a renormalization $\mathcal{R}$ is a family of transformations 
$\mathcal{R}=\left\{ \mathcal{R}_{p}\right\} _{p>0}$ -- mapping probability
distributions to probability distributions -- which is \emph{consistent}: A $%
p$-renormalization followed by a $q$-renormalization equals a $pq$%
-renormalization: $\mathcal{R}_{p}\circ \mathcal{R}_{q}=\mathcal{R}_{pq}$ ($%
p,q>0$; the sign $\circ $ denoting composition).

A probability distribution $D$ is a \emph{fixed point} of the
renormalization $\mathcal{R}$ if it is a fixed point of each of the
renormalization's transformations: $\mathcal{R}_{p}(D)=D$ (for all $p>0$).
The connection between $\mathcal{C}$-fractal probability distributions and
renormalizations is given by the following definition:

\begin{definition}
A renormalization $\mathcal{R}$ is $\mathcal{C}$-fractal if its set of fixed
points coincides with the set of $\mathcal{C}$-fractal probability
distributions.
\end{definition}

In this section we study $\mathcal{C}$-fractality with regard to each of the
distribution-characteristics specified above. As shall be demonstrated,
different distribution-characteristics will lead to very different meanings
of fractality.

\subsection{\label{2.1}Fractality via Frequencies}

Pareto's approach to analyzing the empirical data he gathered was based on 
\emph{frequencies}: studying the occurrence-frequencies of the different
population-values. In other words, Pareto focused on the Probability Density
Function (PDF) $f_{D}(\cdot )$ of a given probability distribution $D$.

The PDF $f_{D}(\cdot )$, in turn, induces the two most fundamental
distribution-characteristics of a probability distribution $D$: \textbf{(i)}
the Cumulative Distribution Function (CDF) $F_{D}(\cdot )$, given by 
\begin{equation}
F_{D}(\theta )=\int_{0}^{\theta }f_{D}(x)dx  \label{21.01}
\end{equation}
($\theta >0$); and, \textbf{(ii)} the Survival Distribution Function (SDF) $%
\overline{F}_{D}(\cdot )$, given by 
\begin{equation}
\overline{F}_{D}(\theta )=\int_{\theta }^{\infty }f(x)dx  \label{21.02}
\end{equation}
($\theta >0$).

In this Subsection we study CDF-fractality and SDF-fractality.

\subsubsection{CDF-fractality}

The CDF $F_{D}(\cdot )$ is monotone increasing from the level $\lim_{\theta
\rightarrow 0}F_{D}(\theta )=0$ to the level $\lim_{\theta \rightarrow
\infty }F_{D}(\theta )=1$. Hence, in order that the CDF $F_{D}(\cdot )$
admit a power-law structure its underlying probability distribution $D$ must
be \emph{bounded from above}. Admissible power-law CDFs are thus of the form 
\begin{equation}
F_{D}(\theta )=\left( \frac{\theta }{a}\right) ^{\alpha }  \label{21.11}
\end{equation}
($0<\theta <a$), where the upper bound $a$ and the exponent $\alpha $ are
arbitrary positive parameters.

With no loss of generality, the upper bound can be set to unity ($a=1$) --
yielding the \emph{Beta }CDFs: 
\begin{equation}
F_{D}(\theta )=\theta ^{\alpha }  \label{21.12}
\end{equation}
($0<\theta <1$).

\bigskip

Let $\xi $ denote a random variable drawn from an arbitrary probability
distribution $D$ supported on the unit interval $(0,1)$. The \emph{%
conditional} distribution of the \emph{scaled} random variable $\xi /p$ --
contingent on the information that the random variable $\xi $ is \emph{no
greater} than the level $p$ -- is given by 
\begin{equation}
\text{\textbf{Prob}}\left( \frac{\xi }{p}\leq \theta \text{ \TEXTsymbol{\vert%
} }\xi \leq p\right) =\frac{\text{\textbf{Prob}}\left( \xi \leq p\theta
\right) }{\text{\textbf{Prob}}\left( \xi \leq p\right) }  \label{21.13}
\end{equation}%
($0<p,\theta <1$). The conditional distribution of equation (\ref{21.13})
induces the conditional renormalization 
\begin{equation}
\left( \overset{\text{ }}{\mathcal{R}_{p}(F_{D})}\right) (\theta )=\frac{%
F_{D}(p\theta )}{F_{D}(p)}  \label{21.14}
\end{equation}%
($0<p,\theta <1$).\footnote{%
In this case the renormalization parameter $p$ is restricted to the range $%
0<p<1$.}

A CDF $F_{D}(\cdot )$ is thus a renormalization fixed point if and only if
it satisfies the functional equation $F_{D}(xy)=F_{D}(x)F_{D}(y)$ ($0<x,y<1$%
). The solutions of this functional equation, in turn, are the \emph{Beta }%
CDFs of equation (\ref{21.12}).

\bigskip

For probability distributions supported on the unit interval $(0,1)$ we
conclude that:

\begin{enumerate}
\item[$\bullet $] The CDF-fractal probability distributions are the \emph{%
Beta distributions} of equation (\ref{21.12}).

\item[$\bullet $] The CDF-fractal renormalization is the \emph{conditional
renormalization} of equation (\ref{21.14}).
\end{enumerate}

\subsubsection{SDF-fractality}

The SDF $\overline{F}_{D}(\cdot )$ is monotone decreasing from the level $%
\lim_{\theta \rightarrow 0}\overline{F}_{D}(\theta )=1$ to the level $%
\lim_{\theta \rightarrow \infty }\overline{F}_{D}(\theta )=0$. Hence, in
order that the SDF $\overline{F}_{D}(\cdot )$ admit a power-law structure
its underlying probability distribution $D$ must be \emph{bounded from below}%
. Admissible power-law SDFs are thus of the form 
\begin{equation}
\overline{F}_{D}(\theta )=\left( \frac{a}{\theta }\right) ^{\alpha }
\label{21.21}
\end{equation}
($\theta >a$), where the lower bound $a$ and the exponent $\alpha $ are
arbitrary positive parameters.

With no loss of generality, the lower bound can be set to unity ($a=1$) --
yielding the \emph{Pareto }SDFs: 
\begin{equation}
\overline{F}_{D}(\theta )=\theta ^{-\alpha }  \label{21.22}
\end{equation}
($\theta >1$).

\bigskip

Let $\xi $ denote a random variable drawn from an arbitrary probability
distribution $D$ supported on the ray $(1,\infty )$. The \emph{conditional}
distribution of the \emph{scaled} random variable $\xi /p$ -- contingent on
the information that the random variable $\xi $ is \emph{greater} than the
level $p$ -- is given by 
\begin{equation}
\text{\textbf{Prob}}\left( \frac{\xi }{p}>\theta \text{ \TEXTsymbol{\vert} }%
\xi >p\right) =\frac{\text{\textbf{Prob}}\left( \xi >p\theta \right) }{\text{%
\textbf{Prob}}\left( \xi >p\right) }  \label{21.23}
\end{equation}
($p,\theta >1$). The conditional distribution of equation (\ref{21.23})
induces the conditional renormalization 
\begin{equation}
\left( \overset{\text{ }}{\mathcal{R}_{p}(\overline{F}_{D})}\right) (\theta
)=\frac{\overline{F}_{D}(p\theta )}{\overline{F}_{D}(p)}  \label{21.24}
\end{equation}
($p,\theta >1$).\footnote{%
In this case the renormalization parameter $p$ is restricted to the range $%
p>1$.}

A SDF $\overline{F}_{D}(\cdot )$ is thus a renormalization fixed point if
and only if it satisfies the functional equation $\overline{F}_{D}(xy)=%
\overline{F}_{D}(x)\overline{F}_{D}(y)$ ($x,y>1$). The solutions of this
functional equation, in turn, are the \emph{Pareto }SDFs of equation (\ref%
{21.22}).

\bigskip

For probability distributions supported on the ray $(1,\infty )$ we conclude
that:

\begin{enumerate}
\item[$\bullet $] The SDF-fractal probability distributions are the \emph{%
Pareto distributions} of equation (\ref{21.22}).

\item[$\bullet $] The SDF-fractal renormalization is the \emph{conditional
renormalization} of equation (\ref{21.24}).
\end{enumerate}

\subsubsection{Exponential representations}

Both the aforementioned \emph{Beta} and the \emph{Pareto} probability
distributions posses an underlying \emph{Exponential structure} which we now
describe.

Let $\xi _{\text{Beta}}$ denote a random variable governed by the \emph{Beta}
CDF of equation (\ref{21.11}); let $\xi _{\text{Pareto}}$ denote a random
variable governed by the \emph{Pareto} SDF of equation (\ref{21.21}); and,
let $\mathcal{E}$ denote an Exponentially-distributed random variable with
unit mean. It is straightforward to observe that the following exponential
representations holds: 
\begin{equation}
\xi _{\text{Beta}}\overset{\text{Law}}{=}a\exp \left\{ -\frac{1}{\alpha }%
\mathcal{E}\right\} \text{ \ \ \ \ and \ \ \ \ }\xi _{\text{Pareto}}\overset{%
\text{Law}}{=}a\exp \left\{ \frac{1}{\alpha }\mathcal{E}\right\} \text{ .}
\label{21.31}
\end{equation}%
Note that equation (\ref{21.31}) immediately implies a reciprocal connection
between the \emph{Beta} and the \emph{Pareto} random variables: 
\begin{equation}
\xi _{\text{Beta}}\overset{\text{Law}}{=}\frac{1}{\xi _{\text{Pareto}}}\text{
\ \ \ \ and \ \ \ \ }\xi _{\text{Pareto}}\overset{\text{Law}}{=}\frac{1}{\xi
_{\text{Beta}}}\text{ .}  \label{21.32}
\end{equation}

\bigskip

Let $\xi $ denote a random variable drawn from an arbitrary positive-valued
probability distribution $D$. The \emph{conditional} distribution of the 
\emph{translated} random variable $\xi -p$ -- contingent on the information
that the random variable $\xi $ is \emph{greater} than the level $p$ -- is
given by 
\begin{equation}
\text{\textbf{Prob}}\left( \overset{\text{ }}{\xi -p>\theta \text{ 
\TEXTsymbol{\vert} }\xi >p}\right) =\frac{\text{\textbf{Prob}}\left( \xi
>p+\theta \right) }{\text{\textbf{Prob}}\left( \xi >p\right) }  \label{21.33}
\end{equation}
($p,\theta >0$). The conditional distribution of equation (\ref{21.33})
induces the conditional renormalization 
\begin{equation}
\left( \overset{\text{ }}{\mathcal{R}_{p}(\overline{F}_{D})}\right) (\theta
)=\frac{\overline{F}_{D}(p+\theta )}{\overline{F}_{D}(p)}  \label{21.34}
\end{equation}
($p,\theta >1$).

Equations (\ref{21.33})-(\ref{21.34}) are the \emph{translational}
counterparts of equations (\ref{21.13})-(\ref{21.14}) and equations (\ref%
{21.23})-(\ref{21.24}).

A SDF $\overline{F}_{D}(\cdot )$ is a renormalization fixed point of
equation (\ref{21.34}) if and only if it satisfies the functional equation $%
\overline{F}_{D}(x+y)=\overline{F}_{D}(x)\overline{F}_{D}(y)$ ($x,y>0$). The
unique unit-mean solution of this functional equation is the unit-mean \emph{%
Exponential} SDF. This characterizing property of the \emph{Exponential
distribution} -- often referred to as \textquotedblleft lack of
memory\textquotedblright\ (\cite{Fel1}, Section XVII.6) -- is of prime
importance in probability theory and its applications. As we see here, this
elemental property also underlies the CDF-fractal renormalization and the
SDF-fractal renormalization.

\subsection{\label{2.2}Fractality via Hazard Rates}

Let $\xi $ denote a random variable drawn from an arbitrary positive-valued
probability distribution $D$. What is the probability that the random
variable $\xi $ be realized \emph{at} the level $\theta $ -- provided that
it is \emph{not} realized \emph{above} the level $\theta $? The answer to
this question is given by the Backward Hazard Rate (BHR) $H_{D}(\cdot )$,
defined as follows: 
\begin{equation}
H_{D}(\theta )=\lim_{\delta \rightarrow 0}\frac{1}{\delta }\mathbf{P}\left( 
\overset{\text{ }}{\xi >\theta -\delta \text{ \TEXTsymbol{\vert} }\xi \leq
\theta }\right) =\frac{f_{D}(\theta )}{F_{D}(\theta )}  \label{22.01}
\end{equation}
($\theta >0$).

And what about the probability that the random variable $\xi $ be realized 
\emph{at} the level $\theta $ -- provided that it is \emph{not} realized 
\emph{below} the level $\theta $? The answer to this analogous question is
given by the Forward Hazard Rate (FHR) $\overline{H}_{D}(\cdot )$, defined
as follows: 
\begin{equation}
\overline{H}_{D}(\theta )=\lim_{\delta \rightarrow 0}\frac{1}{\delta }%
\mathbf{P}\left( \overset{\text{ }}{\xi \leq \theta +\delta \text{ 
\TEXTsymbol{\vert} }\xi >\theta }\right) =\frac{f(\theta )}{\overline{F}%
_{D}(\theta )}  \label{22.02}
\end{equation}
($\theta >0$). The FHR plays a central role in Applied Probability and in
the Theory of Reliability \cite{BP}-\cite{Ros1}.

Both the BHR and the FHR are distribution-characteristics. Indeed, the CDF
and the SDF can be reconstructed, respectively, from the BHR and the FHR via 
\begin{equation}
F_{D}(\theta )=\exp \left\{ -\int_{\theta }^{\infty }H_{D}(x)dx\right\}
\label{22.03}
\end{equation}
($\theta >0$), and via 
\begin{equation}
\overline{F}_{D}(\theta )=\exp \left\{ -\int_{0}^{\theta }\overline{H}%
_{D}(x)dx\right\}  \label{22.04}
\end{equation}
($\theta >0$).

In this subsection we study BHR-fractality and FHR-fractality.

\subsubsection{BHR-fractality}

As indicated above, the CDF $F_{D}(\cdot )$ is monotone increasing from the
level $\lim_{\theta \rightarrow 0}F_{D}(\theta )=0$ to the level $%
\lim_{\theta \rightarrow \infty }F_{D}(\theta )=1$. Hence, equation (\ref%
{22.03}) implies that the BHR $H_{D}(\cdot )$ is integrable at infinity, and
is non-integrable over the entire positive half-line ($\int_{0}^{\infty
}H_{D}(x)dx=\infty $).

Admissible power-law BHRs thus yield the \emph{Fr\'{e}chet} CDFs 
\begin{equation}
F_{D}(\theta )=\exp \left\{ -a\theta ^{-\alpha }\right\}  \label{22.11}
\end{equation}
($\theta >0$), where the coefficient $a$ and the exponent $\alpha $ are
arbitrary positive parameters.

\bigskip

Let $\left\{ \xi _{1},\cdots ,\xi _{n}\right\} $ denote a sequence of $n$
IID random variables drawn from an arbitrary positive-valued probability
distribution $D$. The distribution of the \emph{maximal} random variable $%
\max \left\{ \xi _{1},\cdots ,\xi _{n}\right\} $ -- \emph{scaled-down} by
the multiplicative factor $n^{-1/\alpha }$ -- is given by 
\begin{equation}
\text{\textbf{Prob}}\left( \frac{1}{n^{1/\alpha }}\max \left\{ \xi
_{1},\cdots ,\xi _{n}\right\} \leq \theta \right) =\left( \text{\textbf{Prob}%
}\left( \overset{\text{ }}{\xi _{1}\leq n^{1/\alpha }\theta }\right) \right)
^{n}  \label{22.12}
\end{equation}
($\theta >0$). The maximum distribution of equation (\ref{22.12}) induces
the maximal renormalization 
\begin{equation}
\left( \overset{\text{ }}{\mathcal{R}_{p}(F_{D})}\right) (\theta )=\left( 
\overset{\text{ }}{F_{D}\left( p^{1/\alpha }\theta \right) }\right) ^{p}
\label{22.13}
\end{equation}
($p,\theta >0$).

A CDF $F_{D}(\cdot )$ is thus a renormalization fixed point if and only if
its logarithm $G_{D}(\cdot )=\ln \left( F_{D}(\cdot )\right) $ satisfies the
functional equation $G_{D}(xy)=x^{-\alpha }G_{D}(y)$ ($x,y>0$). The
solutions of this functional equation, in turn, are the \emph{Fr\'{e}chet}
CDFs of equation (\ref{22.11}).

\bigskip

For positive-valued probability distributions we conclude that:

\begin{enumerate}
\item[$\bullet $] The BHR-fractal probability distributions are the \emph{Fr%
\'{e}chet distributions} of equation (\ref{22.11}).

\item[$\bullet $] The BHR-fractal renormalization is the \emph{maximal
renormalization} of equation (\ref{22.13}).
\end{enumerate}

\subsubsection{FHR-fractality}

As indicated above, the SDF $\overline{F}_{D}(\cdot )$ is monotone
decreasing from the level $\lim_{\theta \rightarrow 0}\overline{F}%
_{D}(\theta )=1$ to the level $\lim_{\theta \rightarrow \infty }\overline{F}%
_{D}(\theta )=0$. Hence, equation (\ref{22.04}) implies that the FHR $%
\overline{H}_{D}(\cdot )$ is integrable at the origin, and is non-integrable
over the entire positive half-line ($\int_{0}^{\infty }\overline{H}%
_{D}(x)dx=\infty $).

Admissible power-law FHRs thus yield the \emph{Weibull} SDFs 
\begin{equation}
\overline{F}_{D}(\theta )=\exp \left\{ -a\theta ^{\alpha }\right\}
\label{22.21}
\end{equation}
($\theta >0$), where the coefficient $a$ and the exponent $\alpha $ are
arbitrary positive parameters.

\bigskip

Let $\left\{ \xi _{1},\cdots ,\xi _{n}\right\} $ denote a sequence of $n$
IID random variables drawn from an arbitrary positive-valued probability
distribution $D$. The distribution of the \emph{minimal} random variable $%
\min \left\{ \xi _{1},\cdots ,\xi _{n}\right\} $ -- \emph{scaled-up} by the
multiplicative factor $n^{1/\alpha }$ -- is given by 
\begin{equation}
\text{\textbf{Prob}}\left( \overset{\text{ }}{n^{1/\alpha }\cdot \min
\left\{ \xi _{1},\cdots ,\xi _{n}\right\} >\theta }\right) =\left( \text{%
\textbf{Prob}}\left( \xi _{1}>\frac{\theta }{n^{1/\alpha }}\right) \right)
^{n}  \label{22.22}
\end{equation}
($\theta >0$). The minimum distribution of equation (\ref{22.22}) induces
the minimal renormalization 
\begin{equation}
\left( \overset{\text{ }}{\mathcal{R}_{p}(\overline{F}_{D})}\right) (\theta
)=\left( \overline{F}_{D}\left( \frac{\theta }{p^{1/\alpha }}\right) \right)
^{p}  \label{22.23}
\end{equation}
($p,\theta >0$).

A SDF $\overline{F}_{D}(\cdot )$ is thus a renormalization fixed point if
and only if its logarithm $\overline{G}_{D}(\cdot )=\ln \left( \overline{F}%
_{D}(\cdot )\right) $ satisfies the functional equation $\overline{G}%
_{D}(xy)=x^{\alpha }\overline{G}_{D}(y)$ ($x,y>0$). The solutions of the
this functional equation, in turn, are the \emph{Weibull} SDFs of equation (%
\ref{22.21}).

\bigskip

For positive-valued probability distributions we conclude that:

\begin{enumerate}
\item[$\bullet $] The FHR-fractal probability distributions are the \emph{%
Weibull distributions} of equation (\ref{22.21}).

\item[$\bullet $] The FHR-fractal renormalization is the \emph{minimal
renormalization} of equation (\ref{22.23}).
\end{enumerate}

\subsubsection{Exponential representations}

Both the aforementioned \emph{Fr\'{e}chet} and the \emph{Weibull}
probability distributions posses an underlying \emph{Exponential structure}
which we now describe.

Let $\xi _{\text{Fr\'{e}chet}}$ denote a random variable governed by the 
\emph{Fr\'{e}chet} CDF of equation (\ref{22.11}); let $\xi _{\text{Weibull}}$
denote a random variable governed by the \emph{Weibull} SDF of equation (\ref%
{22.21}); and, let $\mathcal{E}$ denote an Exponentially-distributed random
variable with unit mean. It is straightforward to observe that the following
exponential representations hold: 
\begin{equation}
\xi _{\text{Fr\'{e}chet}}\overset{\text{Law}}{=}\left( \frac{1}{a}\mathcal{E}%
\right) ^{1/\alpha }\text{ \ \ \ \ and \ \ \ \ }\xi _{\text{Weibull}}\overset%
{\text{Law}}{=}\left( \frac{1}{a}\mathcal{E}\right) ^{-1/\alpha }\text{ .}
\label{22.31}
\end{equation}%
Note that equation (\ref{22.31}) immediately implies a reciprocal connection
between the \emph{Fr\'{e}chet} and the \emph{Weibull} random variables: 
\begin{equation}
\xi _{\text{Fr\'{e}chet}}\overset{\text{Law}}{=}\frac{1}{\xi _{\text{Weibull}%
}}\text{ \ \ \ \ and \ \ \ \ }\xi _{\text{Weibull}}\overset{\text{Law}}{=}%
\frac{1}{\xi _{\text{Fr\'{e}chet}}}\text{ .}  \label{22.32}
\end{equation}

The \emph{Exponential distribution} corresponds to the \emph{minimal
renormalization} of equation (\ref{22.23}) with exponent $\alpha =1$.
Indeed, the unique unit-mean solution of this renormalization is the
unit-mean \emph{Exponential} SDF.

\subsection{\label{2.3}Fractality via Laplace-space characteristics}

So forth, we considered fractality via \textquotedblleft
frequency-based\textquotedblright\ distribution-characteristics: CDFs, SDFs,
BHRs, FHRs. In this Subsection we shift to \emph{Laplace space} and turn to
study fractality via the following \textquotedblleft
analytic-based\textquotedblright\ distribution-characteristics: Laplace
Transforms; Moment Sequences; Log-Laplace Transforms; Cumulant Sequences.

\subsubsection{Laplace Transforms and Moment Sequences}

The Laplace Transform (LT) $L_{D}(\cdot )$ of a positive-valued probability
distribution $D$ is the Laplace Transform of its PDF $f_{D}(\cdot )$: 
\begin{equation}
L_{D}(\theta )=\int_{0}^{\infty }\exp \left\{ -\theta x\right\} f_{D}(x)dx
\label{23.11}
\end{equation}
($\theta \geq 0$). The LT is a distribution-characteristic -- though, in
general, the reconstruction of a PDF from a given LT is hard a task \cite%
{Wid}.

The LT $L_{D}(\cdot )$ is monotone decreasing from the level $L_{D}(0)=1$ to
the level $\lim_{\theta \rightarrow \infty }L_{D}(\theta )=0$. Hence, LTs 
\emph{cannot} admit the power-law structure of equation (\ref{2.01}).

\bigskip

In case the LT $L_{D}(\cdot )$ admits a Taylor expansion around the origin,
the Moment Sequence (MS) $\left\{ M_{D}(n)\right\} _{n=0}^{\infty }$ of the
probability distribution $D$ is well defined and is given by 
\begin{equation}
L_{D}(\theta )=\sum_{n=0}^{\infty }M_{D}(n)\frac{(-\theta )^{n}}{n!}
\label{23.21}
\end{equation}
($\theta \geq 0$). Reconstructing a probability distribution from a given MS
is known as the \emph{Stieltjes Moment Problem} \cite{Wid}.

Since the Moment of order zero equals unity ($M_{D}(0)=1$) MSs \emph{cannot}
admit the power-law structure of equation (\ref{2.01}).

\bigskip

We conclude that there are \emph{no} LT-fractal and \emph{no} MS-fractal
probability distributions.

\subsubsection{Log-Laplace Transforms}

A ``cousin'' of the LT $L_{D}(\cdot )$ is its logarithm -- referred to as
the Log-Laplace Transform (LLT) $\Psi _{D}(\cdot )$ and given by 
\begin{equation}
\Psi _{D}(\theta )=-\ln \left( L_{D}(\theta )\right)  \label{23.31}
\end{equation}
($\theta \geq 0$).

The LLT initiates at the origin ($\Psi _{D}(0)=0$) and is monotone
increasing ($\Psi _{D}^{\prime }(\theta )>0$) and concave ($\Psi
_{D}^{\prime }(\theta )<0$).

Admissible power-law LLTs are thus the \emph{L\'{e}vy Stable} LLTs: 
\begin{equation}
\Psi _{D}(\theta )=a\theta ^{\alpha }  \label{23.32}
\end{equation}
($\theta \geq 0$), where $a$ is an arbitrary positive coefficient and where
the exponent $\alpha $ takes values in the range $0<\alpha <1$. The \emph{L%
\'{e}vy Stable} LLTs of equation (\ref{23.32}) admit the integral
representation 
\begin{equation}
\Psi _{D}(\theta )=\theta \int_{0}^{\infty }\exp \left\{ -\theta x\right\}
\left( \frac{a}{\Gamma (1-\alpha )}\frac{1}{x^{\alpha }}\right) dx
\label{23.33}
\end{equation}
($\theta \geq 0$) -- whose meaning will be explained in the sequel.

(Apart from the special case $\alpha =1/2$, there is no ``closed form''
representation for the PDFs of the \emph{L\'{e}vy Stable }probability
distributions.)

\bigskip

Let $\left\{ \xi _{1},\cdots ,\xi _{n}\right\} $ denote a sequence of $n$
IID random variables drawn from an arbitrary positive-valued probability
distribution $D$. The LT of the \emph{aggregate} $\xi _{1}+\cdots +\xi _{n}$
-- \emph{scaled-down} by the multiplicative factor $n^{-1/\alpha }$ -- is
given by 
\begin{equation}
\left\langle \exp \left\{ -\theta \frac{\xi _{1}+\cdots +\xi _{n}}{%
n^{1/\alpha }}\right\} \right\rangle =\left\langle \exp \left\{ -\frac{%
\theta }{n^{1/\alpha }}\xi _{1}\right\} \right\rangle ^{n}  \label{23.34}
\end{equation}
($\theta \geq 0$). The LT of equation (\ref{23.34}) induces the aggregative
renormalization 
\begin{equation}
\left( \overset{\text{ }}{\mathcal{R}_{p}(\Psi _{D})}\right) (\theta )=p\Psi
_{D}\left( \frac{\theta }{p^{1/\alpha }}\right)  \label{23.35}
\end{equation}
($p>0$, $\theta \geq 0$).

A LLT $\Psi _{D}(\cdot )$ is thus a renormalization fixed point if and only
if it satisfies the functional equation $\Psi _{D}(xy)=x^{-\alpha }\Psi
_{D}(y)$ ($x,y>0$). The solutions of this functional equation, in turn, are
the \emph{L\'{e}vy Stable} LLTs of equation (\ref{23.32}).

\bigskip

For positive-valued probability distributions we conclude that:

\begin{enumerate}
\item[$\bullet $] The LLT-fractal probability distributions are the \emph{L%
\'{e}vy Stable distributions} characterized by the LLTs of equation (\ref%
{23.32}).

\item[$\bullet $] The LLT-fractal renormalization is the \emph{aggregative
renormalization} of equation (\ref{23.35}).
\end{enumerate}

\subsubsection{Cumulant Sequences}

In case the LLT $\Psi _{D}(\cdot )$ admits a Taylor expansion around the
origin, the Cumulant Sequence (CS) $\left\{ C_{D}(n)\right\} _{n=1}^{\infty
} $ of the probability distribution $D$ is well defined and is given by 
\begin{equation}
\Psi _{D}(\theta )=-\sum_{n=1}^{\infty }C_{D}(n)\frac{(-\theta )^{n}}{n!}
\label{23.41}
\end{equation}
($\theta \geq 0$).

Power-law CSs of the form $C_{D}(n)=an^{-\alpha }$ ($n=1,2,\cdots $), where
the coefficient $a$ and the exponent $\alpha $ are arbitrary positive
parameters, yield LLTs admitting the following integral representation: 
\begin{equation}
\Psi _{D}(\theta )=\theta \int_{0}^{1}\exp \left\{ -\theta x\right\} \left( 
\frac{a}{\Gamma (1+\alpha )}\left( -\ln (x)\right) ^{\alpha }\right) dx
\label{23.42}
\end{equation}
($\theta \geq 0$).

The proof of equation (\ref{23.42}) is given in the Appendix; the meaning of
this integral representation will be explained in the sequel. The
renormalization associated with CS-fractal probability distributions is
based on their underlying Poissonian structure -- which, too, will be
explained in the sequel.

\bigskip

For positive-valued probability distributions we conclude that:

\begin{enumerate}
\item[$\bullet $] The CS-fractal probability distributions are characterized
by the LLTs of equation (\ref{23.42}).

\item[$\bullet $] The CS-fractal renormalization is a \emph{Poissonian
renormalization} (yet to be presented).
\end{enumerate}

\subsection{\label{2.4}Interim summery}

Table \emph{1} summarizes the six classes of fractal probability
distributions characterized in this Section.

We note that in the context of IID sequences of positive-valued random
variables: \textbf{(i)} Extreme Value Theory asserts that the \emph{Fr\'{e}%
chet} and \emph{Weibull} distributions are, respectively, the only possible
linear scaling limits of the sequences' maxima and minima \cite{Gne}-\cite%
{Gal}; \textbf{(ii)} the Central Limit Theorem asserts that the one-sided 
\emph{L\'{e}vy Stable} distribution is the only possible linear scaling
limit of the sequences' sums \cite{Lev}-\cite{IL}.

\begin{center}
{\large Table 1}

\textbf{The classes of fractal probability distributions}

\bigskip

\begin{tabular}{||c||c||c||}
\hline\hline
$%
\begin{array}{c}
\text{\textbf{\ }} \\ 
\text{\textbf{Fractality}} \\ 
\text{\textbf{\ }}%
\end{array}
$ & $%
\begin{array}{c}
\text{\textbf{\ }} \\ 
\text{\textbf{Distribution}} \\ 
\text{\textbf{\ }}%
\end{array}
$ & $%
\begin{array}{c}
\text{\textbf{\ }} \\ 
\text{\textbf{Renormalization}} \\ 
\text{\textbf{\ }}%
\end{array}
$ \\ \hline\hline
$%
\begin{array}{c}
\text{ } \\ 
\text{CDF-fractal} \\ 
\text{ }%
\end{array}
$ & $%
\begin{array}{c}
\text{ } \\ 
\text{\emph{Beta}} \\ 
\text{ }%
\end{array}
$ & $%
\begin{array}{c}
\text{ } \\ 
\text{Conditional} \\ 
\text{ }%
\end{array}
$ \\ \hline\hline
$%
\begin{array}{c}
\text{ } \\ 
\text{SDF-fractal} \\ 
\text{ }%
\end{array}
$ & $%
\begin{array}{c}
\text{ } \\ 
\text{\emph{Pareto}} \\ 
\text{ }%
\end{array}
$ & $%
\begin{array}{c}
\text{ } \\ 
\text{Conditional} \\ 
\text{ }%
\end{array}
$ \\ \hline\hline
$%
\begin{array}{c}
\text{ } \\ 
\text{BHR-fractal} \\ 
\text{ }%
\end{array}
$ & $%
\begin{array}{c}
\text{ } \\ 
\text{\emph{Fr\'{e}chet}} \\ 
\text{ }%
\end{array}
$ & $%
\begin{array}{c}
\text{ } \\ 
\text{Maximal} \\ 
\text{ }%
\end{array}
$ \\ \hline\hline
$%
\begin{array}{c}
\text{ } \\ 
\text{FHR-fractal} \\ 
\text{ }%
\end{array}
$ & $%
\begin{array}{c}
\text{ } \\ 
\text{\emph{Weibull}} \\ 
\text{ }%
\end{array}
$ & $%
\begin{array}{c}
\text{ } \\ 
\text{Minimal} \\ 
\text{ }%
\end{array}
$ \\ \hline\hline
$%
\begin{array}{c}
\text{ } \\ 
\text{LLT-fractal} \\ 
\text{ }%
\end{array}
$ & $%
\begin{array}{c}
\text{ } \\ 
\text{\emph{L\'{e}vy Stable}} \\ 
\text{ }%
\end{array}
$ & $%
\begin{array}{c}
\text{ } \\ 
\text{Aggregative} \\ 
\text{ }%
\end{array}
$ \\ \hline\hline
$%
\begin{array}{c}
\text{ } \\ 
\text{CS-fractal} \\ 
\text{ }%
\end{array}
$ & $%
\begin{array}{c}
\text{ } \\ 
\text{\emph{---}} \\ 
\text{ }%
\end{array}
$ & $%
\begin{array}{c}
\text{ } \\ 
\text{Poissonian} \\ 
\text{ }%
\end{array}
$ \\ \hline\hline
\end{tabular}
\end{center}

\section{\label{3}Poissonian representation of fractal distributions}

In this section we provide \emph{Poissonian representations} for all six
classes of fractal probability distributions characterized in the previous
Section.

\subsection{\label{3.0}Poisson processes}

In this Subsection we recall the notion of Poisson processes. For further
details the readers are referred to \cite{Kin}.

\bigskip

A \emph{Poisson process} $\mathcal{X}$ on the positive half-line, with \emph{%
rate function} $r(\cdot )$, is a random collection of positive-valued points
satisfying the following properties: \textbf{(i)} the number of points $N_{%
\mathcal{X}}(I)$ residing in the interval $I$ is a Poisson-distributed
random variable with mean $\int_{I}r(x)dx$; and, \textbf{(ii)} if $\left\{
I_{k}\right\} _{k}$ is a finite collection of disjoint intervals then $%
\left\{ N_{\mathcal{X}}(I_{k})\right\} _{k}$ is a finite collection of
independent random variables.

The rate function $r(\cdot )$ is the Poissonian analogue of the PDF in the
context of probability distributions. The Poissonian analogues of the CDF
and the SDF, respectively, are: \textbf{(i)} the Cumulative Rate Function
(CRF) $R(\cdot )$, given by 
\begin{equation}
R(\theta )=\int_{0}^{\theta }r(x)dx  \label{30.01}
\end{equation}
($\theta >0$); and, \textbf{(ii)} the Survival Rate Function (SRF) $%
\overline{R}(\cdot )$, given by 
\begin{equation}
\overline{R}(\theta )=\int_{\theta }^{\infty }r(x)dx  \label{30.02}
\end{equation}
($\theta >0$).

The CRF $R(\cdot )$ is well defined if and only if the rate function $%
r(\cdot )$ is integrable at the origin -- in which case it is a monotone
non-decreasing function initiating from the origin ($R(0)=0$). The SRF $%
\overline{R}(\cdot )$ is well defined if and only if the rate function $%
r(\cdot )$ is integrable at infinity -- in which case it is a monotone
non-increasing function decreasing to zero ($\lim_{\theta \rightarrow \infty
}R(\theta )=0$).

The \emph{average} number of points of the Poisson process $\mathcal{X}$
residing below the level $\theta $ is given by the CRF value $R(\theta )$;
the \emph{average} number of points residing above the level $\theta $ is
given by the SRF value $\overline{R}(\theta )$.

\subsection{\label{3.1}Poissonian maxima}

Consider the \emph{maximum} of the Poisson process $\mathcal{X}$, defined as
follows: 
\begin{equation}
X_{\text{max}}=\max_{x\in \mathcal{X}}\left\{ x\right\} \text{ .}
\label{31.01}
\end{equation}
We refer to the probability distribution of the random variable $X_{\text{max%
}}$ as the \emph{maximal distribution} of the Poisson process $\mathcal{X}$.

The maximum $X_{\text{max}}$ is smaller than the level $\theta $ ($\theta >0$%
) if and only if the process $\mathcal{X}$ has no points residing above this
level. Namely: $\left\{ X_{\text{max}}\leq \theta \right\} =\left\{ N_{%
\mathcal{X}}((\theta ,\infty ))=0\right\} $. Since the random variable $N_{%
\mathcal{X}}((\theta ,\infty ))$ is Poisson-distributed with mean $\overline{%
R}(\theta )$, we obtain that the maximal distribution of the Poisson process 
$\mathcal{X}$ is characterized by the CDF 
\begin{equation}
F_{\text{max}}(\theta )=\exp \left\{ -\overline{R}(\theta )\right\}
\label{31.02}
\end{equation}
($\theta >0$).

Equation (\ref{31.02}) implies a one-to-one correspondence between Poisson
processes (characterized by their SRFs $\overline{R}(\cdot )$) and their
associated maximal distributions (characterized by their CDFs $F_{\text{max}%
}(\cdot )$). This one-to-one correspondence yields the following Poissonian
representation of the CDF-fractal and BHR-fractal probability distributions:

\begin{enumerate}
\item[$\bullet $] A probability distribution $D$ is \emph{CDF-fractal} if
and only if it is the maximal distribution of a Poisson process $\mathcal{X}$
with \emph{logarithmic} SRF of the form 
\begin{equation}
\overline{R}(\theta )=-\ln \left( \left( \frac{\theta }{a}\right) ^{\alpha
}\right)  \label{31.03}
\end{equation}
($0<\theta <a$), where the upper bound $a$ and the exponent $\alpha $ are
arbitrary positive parameters.

\item[$\bullet $] A probability distribution $D$ is \emph{BHR-fractal} if
and only if it is the maximal distribution of a Poisson process $\mathcal{X}$
with \emph{power-law} SRF of the form 
\begin{equation}
\overline{R}(\theta )=a\theta ^{-\alpha }  \label{31.04}
\end{equation}
($\theta >0$), where the coefficient $a$ and the exponent $\alpha $ are
arbitrary positive parameters.
\end{enumerate}

Equations (\ref{31.03}) and (\ref{31.04}) follow, respectively, from
equations (\ref{21.11}) and (\ref{22.11}).

\subsection{\label{3.2}Poissonian minima}

Consider the \emph{minimum} of the Poisson process $\mathcal{X}$, defined as
follows: 
\begin{equation}
X_{\text{min}}=\min_{x\in \mathcal{X}}\left\{ x\right\} \text{ .}
\label{32.01}
\end{equation}
We refer to the probability distribution of the random variable $X_{\text{min%
}}$ as the \emph{minimal distribution} of the Poisson process $\mathcal{X}$.

The minimum $X_{\text{min}}$ is larger than the level $\theta $ ($\theta >0$%
) if and only if Poisson process $\mathcal{X}$ has no points residing below
this level. Namely: $\left\{ X_{\text{min}}>\theta \right\} =\left\{ N_{%
\mathcal{X}}((0,\theta ])=0\right\} $. Since the random variable $N_{%
\mathcal{X}}((0,\theta ])$ is Poisson-distributed with mean $R(\theta )$, we
obtain that the minimal distribution of the Poisson process $\mathcal{X}$ is
characterized by the CDF 
\begin{equation}
\overline{F}_{\text{min}}(\theta )=\exp \left\{ -R(\theta )\right\}
\label{32.02}
\end{equation}
($\theta >0$).

Equation (\ref{32.02}) implies a one-to-one correspondence between Poisson
processes (characterized by their CRFs $R(\cdot )$) and their associated
minimal distributions (characterized by their SDFs $\overline{F}_{\text{min}%
}(\cdot )$). This one-to-one correspondence yields the following Poissonian
representation of the SDF-fractal and FHR-fractal probability distributions:

\begin{enumerate}
\item[$\bullet $] A probability distribution $D$ is \emph{SDF-fractal} if
and only if it is the minimal distribution of a Poisson process $\mathcal{X}$
with \emph{logarithmic} CRF of the form 
\begin{equation}
R(\theta )=-\ln \left( \left( \frac{a}{\theta }\right) ^{\alpha }\right)
\label{32.03}
\end{equation}
($\theta >a$), where the lower bound $a$ and the exponent $\alpha $ are
arbitrary positive parameters.

\item[$\bullet $] A probability distribution $D$ is \emph{FHR-fractal} if
and only if it is the minimal distribution of a Poisson process $\mathcal{X}$
with \emph{power-law} CRF of the form 
\begin{equation}
R(\theta )=a\theta ^{\alpha }  \label{32.04}
\end{equation}
($\theta >0$), where the coefficient $a$ and the exponent $\alpha $ are
arbitrary positive parameters.
\end{enumerate}

Equations (\ref{32.03}) and (\ref{32.04}) follow, respectively, from
equations (\ref{21.21}) and (\ref{22.21}).

\subsection{\label{3.3}Poissonian aggregates}

Consider the \emph{aggregate} of the Poisson process $\mathcal{X}$, defined
as follows: 
\begin{equation}
X_{\text{agg}}=\sum_{x\in \mathcal{X}}x\text{ .}  \label{33.01}
\end{equation}
We refer to the probability distribution of the random variable $X_{\text{agg%
}}$ as the \emph{aggregate distribution} of the Poisson process $\mathcal{X}$%
.

The aggregate of equation (\ref{33.01}) can be either convergent ($X_{\text{%
agg}}<\infty $) or divergent ($X_{\text{agg}}=\infty $). Campbell's theorem
of the theory of Poisson processes (\cite{Kin}, Section \emph{3.2}) implies
that the aggregate is convergent if and only if the SRF $\overline{R}(\cdot )
$ is integrable at the origin -- in which case the aggregate distribution of
the Poisson process $\mathcal{X}$ is characterized by the LLT 
\begin{equation}
\Psi _{\text{agg}}(\theta )=\theta \int_{0}^{\infty }\exp \left\{ -\theta
x\right\} \overline{R}(x)dx  \label{33.02}
\end{equation}%
($\theta \geq 0$).

Equation (\ref{33.02}) implies a one-to-one correspondence between Poisson
processes (characterized by their SRFs $\overline{R}(\cdot )$) and their
associated aggregate distributions (characterized by their LLTs $\Psi _{%
\text{agg}}(\cdot )$ -- which, in turn, are characterized by the Laplace
transforms of the underlying SRFs). This one-to-one correspondence yields
the following Poissonian representation of the LLT-fractal and CS-fractal
probability distributions:

\begin{enumerate}
\item[$\bullet $] A probability distribution $D$ is \emph{LLT-fractal} if
and only if it is the aggregate distribution of a Poisson process $\mathcal{X%
}$ with \emph{power-law} SRF of the form 
\begin{equation}
\overline{R}(\theta )=\frac{a}{\Gamma (1-\alpha )}\frac{1}{\theta ^{\alpha }}
\label{33.03}
\end{equation}
($\theta >0$), where $a$ is an arbitrary positive coefficient and where the
exponent $\alpha $ takes values in the range $0<\alpha <1$.

\item[$\bullet $] A probability distribution $D$ is \emph{CS-fractal} if and
only if it is the aggregate distribution of a Poisson process $\mathcal{X}$
with \emph{logarithmic} SRF of the form 
\begin{equation}
\overline{R}(\theta )=\frac{a}{\Gamma (1+\alpha )}\left( \overset{\text{ }}{%
-\ln (\theta )}\right) ^{\alpha }  \label{33.04}
\end{equation}
($0<\theta <1$), where the coefficient $a$ and the exponent $\alpha $ are
arbitrary positive parameters.
\end{enumerate}

Equations (\ref{33.03}) and (\ref{33.04}) follow, respectively, from
equations (\ref{23.33}) and (\ref{23.42}).

\section{\label{4}The underlying Poissonian fractals}

In Section \ref{2} we characterized six classes of fractal probability
distributions -- each stemming from a different distribution-characteristic,
and each associated with a different renormalization. In section \ref{3} we
have further seen that all six classes of fractal probability distributions
admit Poissonian representations -- either maximal, minimal, or aggregative.

\emph{Is there any kind of an underlying order to this ``little zoo'' of
fractal distributions?}

The answer is affirmative: all fractal distributions obtained are \emph{%
functional projections} of underlying \emph{Poissonian fractals} -- as we
shall show in this Section.

\subsection{\label{4.1}Poissonian renormalizations and their fixed points}

In this Subsection we study renormalizations of Poisson processes defined on
the positive half-line. We follow the renormalization approach used in \cite%
{EK1}.

\subsubsection{Poissonian renormalizations}

Let $\left\{ \phi _{p}\right\} _{p>0}$ be a family of \emph{consistent
scaling functions}: monotone-increasing functions which map the positive
half-line $(0,\infty )$ onto itself, and which satisfy the \textquotedblleft
consistency condition\textquotedblright\ $\phi _{p}\circ \phi _{q}=\phi _{pq}
$ ($p,q>0$; the sign $\circ $ denoting composition).

Given a Poisson process $\mathcal{X}$ with rate function $r(\cdot )$ we
construct its $p$\emph{-order renormalization} $\mathcal{X}_{p}$ via the
following two-step algorithm: \textbf{(i) }replace the process $\mathcal{X}$
by an intermediate Poisson process $\mathcal{X}_{p}^{\text{int}}$ with rate
function $r_{p}^{\text{int}}(\cdot )=p\cdot r(\cdot )$;\footnote{%
Note that if $p$ is integer then the intermediate Poisson process $\mathcal{X%
}_{p}^{\text{int}}$ is the \emph{union} of $p$ IID copies of the original
Poisson process $\mathcal{X}$.} \textbf{(ii)} shift the points of the
intermediate process $\mathcal{X}_{p}^{\text{int}}$ using the $p^{\text{th}}$
scaling function $\phi _{p}$. The resulting $p$-order renormalization is
given by 
\begin{equation*}
\mathcal{X}_{p}=\left\{ \phi _{p}(x)\right\} _{x\in \mathcal{X}_{p}^{\text{%
int}}}\text{ \ .}
\end{equation*}

(The ``consistency condition'' is required in order to ensure that the
Poissonian renormalization is consistent. Namely, that a $p$-order
renormalization followed by a $q$-order renormalization equals a $pq$-order
renormalization.)

The connection between the CRF $R_{p}(\cdot )$ and the SRF $\overline{R}%
_{p}(\cdot )$ of the $p$-order renormalization $\mathcal{X}_{p}$, and the
CRF $R(\cdot )$ and the SRF $\overline{R}(\cdot )$ of the original process $%
\mathcal{X}$, is given by: $R_{p}(\cdot )=pR\left( \phi _{p}^{-1}(\cdot
)\right) $ and $\overline{R}_{p}(\cdot )=p\overline{R}\left( \phi
_{p}^{-1}(\cdot )\right) $, where the function $\phi _{p}^{-1}(\cdot )$
denotes the inverse of the scaling function $\phi _{p}(\cdot )$ (these
results are an immediate consequence of the ``displacement theorem'' of the
theory of Poisson processes -- see Section \emph{5.5} in \cite{Kin}).

Denoting by $\mathcal{R}=\left\{ \mathcal{R}_{p}\right\} _{p>0}$ the
Poissonian renormalization defined, we have: 
\begin{equation}
\mathcal{R}_{p}\left( R\right) =p\left( R\circ \phi _{p}^{-1}\right) \text{
\ \ and \ \ }\mathcal{R}_{p}\left( \overline{R}\right) =p\left( \overline{R}%
\circ \phi _{p}^{-1}\right)  \label{41.11}
\end{equation}
($p>0$).

A Poisson process $\mathcal{X}$ is a \emph{fixed point} of the
renormalization $\mathcal{R}$ if it is left statistically unchanged by the
renormalization's action: the $p$-order renormalization $\mathcal{X}_{p}$
being equal, in law, to the original process $\mathcal{X}$. In terms of the
CRF and SRF it is required that $\mathcal{R}_{p}\left( R\right) =R$ and $%
\mathcal{R}_{p}\left( \overline{R}\right) =\overline{R}$ (for all $p>0$).
Using equation (\ref{41.11}) we conclude that: the Poisson process $\mathcal{%
X}$ is a renormalization fixed point if and only if its CRF $R(\cdot )$ and
SRF $\overline{R}(\cdot )$ satisfy 
\begin{equation}
R\circ \phi _{p}=pR\text{ \ \ and \ \ }\overline{R}\circ \phi _{p}=p%
\overline{R}  \label{41.12}
\end{equation}%
($p>0$).

\bigskip

The two most fundamental Poissonian renormalizations are \emph{multiplicative%
} and \emph{power-law}. We now turn to characterize the fixed points of
these renormalizations.

\subsubsection{Fixed points of multiplicative renormalizations}

A \emph{multiplicative renormalization} is based on a set of multiplicative
scaling functions $\left\{ \phi _{p}\right\} _{p>0}$. The consistency
condition implies that the multiplicative scaling functions admit the form 
\begin{equation}
\phi _{p}(x)=p^{\varepsilon }x  \label{41.21}
\end{equation}%
($x>0$), where the exponent $\varepsilon $ is an arbitrary non-zero
parameter.

The fixed points of a multiplicative renormalization governed by the scaling
functions of equation (\ref{41.21}) are as follows (the coefficient $c$
being an arbitrary positive parameter):

\begin{enumerate}
\item[$\bullet $] In the case of a \emph{positive} exponent $\varepsilon $
the renormalization fixed points are characterized by the CRF 
\begin{equation}
R(\theta )=c\theta ^{1/\varepsilon }\text{ \ \ (}\theta >0\text{).}
\label{41.22}
\end{equation}%
The accumulation point of these fixed-point Poisson processes is $x_{\ast
}=\infty $.\footnote{%
A point $x_{\ast }$ ($0\leq x_{\ast }\leq \infty $) is said to be an \emph{%
accumulation point} of the Poisson process $\mathcal{X}$ if, with
probability one, there are infinitely many points of $\mathcal{X}$ within
any given neighborhood of $x_{\ast }$.}

\item[$\bullet $] In the case of a \emph{negative} exponent $\varepsilon $
the renormalization fixed points are characterized by the SRF 
\begin{equation}
\overline{R}(\theta )=c\theta ^{1/\varepsilon }\text{ \ \ (}\theta >0\text{).%
}  \label{41.23}
\end{equation}%
The accumulation point of these fixed-point Poisson processes is $x_{\ast }=0
$.
\end{enumerate}

\subsubsection{Fixed points of power-law renormalizations}

A \emph{power-law renormalization} is based on a set of power-law scaling
functions $\left\{ \phi _{p}\right\} _{p>0}$. The consistency condition
implies that the power-law scaling functions admit the form 
\begin{equation}
\phi _{p}(x)=x^{p^{\varepsilon }}  \label{41.31}
\end{equation}%
($x>0$), where the exponent $\varepsilon $ is an arbitrary non-zero
parameter.

The fixed points of a power-law renormalization governed by the scaling
functions of equation (\ref{41.31}) cannot range over the entire positive
half-line $(0,\infty )$. Rather, they may range either on the unit interval $%
(0,1)$ or on the ray $(1,\infty )$ (note that the power-law scaling
functions of equation (\ref{41.31}) indeed map the unit interval $(0,1)$ and
the ray $(1,\infty )$, respectively, onto themselves).

The fixed points of a power-law renormalization governed by the scaling
functions of equation (\ref{41.31}), and ranging over the unit interval $%
(0,1)$, are as follows (the coefficient $c$ being an arbitrary positive
parameter):

\begin{enumerate}
\item[$\bullet $] In the case of a \emph{positive} exponent $\varepsilon $
the renormalization fixed points are characterized by the SRF 
\begin{equation}
\overline{R}(\theta )=c\left( -\ln \theta \right) ^{1/\varepsilon }\text{ \
\ (}0<\theta <1\text{).}  \label{41.32}
\end{equation}%
The accumulation point of these fixed-point Poisson processes is $x_{\ast }=0
$.

\item[$\bullet $] In the case of a \emph{negative} exponent $\varepsilon $
the renormalization fixed points are characterized by the CRF 
\begin{equation}
R(\theta )=c\left( -\ln \theta \right) ^{1/\varepsilon }\text{ \ \ (}%
0<\theta <1\text{).}  \label{41.33}
\end{equation}%
The accumulation point of these fixed-point Poisson processes is $x_{\ast }=1
$.
\end{enumerate}

The fixed points of a power-law renormalization governed by the scaling
functions of equation (\ref{41.31}), and ranging over the ray $(1,\infty )$,
are as follows (the coefficient $c$ being an arbitrary positive parameter):

\begin{enumerate}
\item[$\bullet $] In the case of a \emph{positive} exponent $\varepsilon $
the renormalization fixed points are characterized by the CRF 
\begin{equation}
R(\theta )=c\left( \ln \theta \right) ^{1/\varepsilon }\text{ \ \ (}\theta >1%
\text{).}  \label{41.34}
\end{equation}%
The accumulation point of these fixed-point Poisson processes is $x_{\ast
}=\infty $.

\item[$\bullet $] In the case of a \emph{negative} exponent $\varepsilon $
the renormalization fixed points are characterized by the SRF 
\begin{equation}
\overline{R}(\theta )=c\left( \ln \theta \right) ^{1/\varepsilon }\text{ \ \
(}\theta >1\text{).}  \label{41.35}
\end{equation}%
The accumulation point of these fixed-point Poisson processes is $x_{\ast }=1
$.
\end{enumerate}

\subsection{\label{4.2}Poissonian fractals}

In the previous Subsection we obtained six classes of renormalization
fixed-point Poisson processes. Excluding the fixed-point processes whose
accumulation point is $x_{\ast }=1$ (an interior point of the positive
half-line), and considering the fixed-point processes whose accumulation
point is either the origin $x_{\ast }=0$ or infinity $x_{\ast }=\infty $
(the boundaries of the positive half-line), we now define four classes of 
\emph{Poissonian fractals}. The fractal distributions of Section \ref{2}
shall turn out to be one-dimensional functional projections -- either
maximal, minimal, or aggregative -- of these underlying Poissonian fractals.

\subsubsection{Linear Poissonian fractals}

The class of \emph{linear Poissonian fractals} comprises of all Poisson
processes governed by CRFs admitting the power-law form 
\begin{equation}
R(\theta )=a\theta ^{\alpha }\text{ \ \ \ \ (}\theta >0\text{),}
\label{42.11}
\end{equation}
where the coefficient $a$ and the exponent $\alpha $ are arbitrary positive
parameters. The members of this class are fixed points of \emph{%
multiplicative} Poissonian renormalizations. For this class:

\begin{enumerate}
\item[$\bullet $] The maximal distribution is degenerate: the maximum $X_{%
\text{max}}$ equals infinity with probability one.

\item[$\bullet $] The minimal distribution is the FHR-fractal \emph{Weibull
distribution}, characterized by the SDF 
\begin{equation}
\overline{F}_{\text{min}}(\theta )=\exp \left\{ -a\theta ^{\alpha }\right\}
\label{42.12}
\end{equation}
($\theta >0$).

\item[$\bullet $] The aggregate distribution is degenerate: the aggregate $%
X_{\text{agg}}$ is infinite with probability one.
\end{enumerate}

\subsubsection{Harmonic Poissonian fractals}

The class of \emph{harmonic Poissonian fractals} comprises of all Poisson
processes governed by SRFs admitting the power-law form 
\begin{equation}
\overline{R}(\theta )=a\theta ^{-\alpha }\text{ \ \ \ \ (}\theta >0\text{),}
\label{42.21}
\end{equation}
where the coefficient $a$ and the exponent $\alpha $ are arbitrary positive
parameters. The members of this class are fixed points of \emph{%
multiplicative} Poissonian renormalizations. For this class:

\begin{enumerate}
\item[$\bullet $] The maximal distribution is the BHR-fractal \emph{Fr\'{e}%
chet distribution}, characterized by the CDF 
\begin{equation}
F_{\text{max}}(\theta )=\exp \left\{ -a\theta ^{-\alpha }\right\}
\label{42.22}
\end{equation}
($\theta >0$).

\item[$\bullet $] The minimal distribution is degenerate: the minimum $X_{%
\text{min}}$ equals zero with probability one.

\item[$\bullet $] If the exponent $\alpha $ is in the range $0<\alpha <1$
then the aggregate distribution is the LLT-fractal \emph{L\'{e}vy Stable
distribution}, characterized by the LLT 
\begin{equation}
\Psi _{\text{agg}}(\theta )=\Gamma (1-\alpha )a\theta ^{\alpha }
\label{42.23}
\end{equation}
($\theta >0$).

\item[$\bullet $] If the exponent $\alpha $ is in the range $\alpha \geq 1$
then the aggregate distribution is degenerate: the aggregate $X_{\text{agg}}$
is infinite with probability one.
\end{enumerate}

\subsubsection{Log-linear Poissonian fractals}

The class of \emph{log-linear Poissonian fractals} comprises of all Poisson
processes governed by CRFs admitting the logarithmic form 
\begin{equation}
R(\theta )=a\left( \ln (\theta )\right) ^{\alpha }\text{ \ \ \ \ (}\theta >1%
\text{),}  \label{42.31}
\end{equation}
where the coefficient $a$ and the exponent $\alpha $ are arbitrary positive
parameters. The members of this class are fixed points of \emph{power-law}
Poissonian renormalizations. For this class:

\begin{enumerate}
\item[$\bullet $] The maximal distribution is degenerate: the maximum $X_{%
\text{max}}$ equals infinity with probability one.

\item[$\bullet $] The minimal distribution is characterized by the SDF 
\begin{equation}
\overline{F}_{\text{min}}(\theta )=\exp \left\{ -a\left( \ln (\theta
)\right) ^{\alpha }\right\}  \label{42.32}
\end{equation}
($\theta >1$).

\item[$\bullet $] The aggregate distribution is degenerate: the aggregate $%
X_{\text{agg}}$ is infinite with probability one.
\end{enumerate}

The SDF of equation (\ref{42.32}) reduces to the \emph{Pareto} SDF when
setting the exponent value $\alpha $ to unity. Thus, we refer to the
distribution corresponding to this SDF as \emph{Hyper Pareto}.

\subsubsection{Log-harmonic Poissonian fractals}

The class of \emph{log-harmonic Poissonian fractals} comprises of all
Poisson processes governed by SRFs admitting the logarithmic form 
\begin{equation}
\overline{R}(\theta )=a\left( -\ln (\theta )\right) ^{\alpha }\text{ \ \ \ \
(}0<\theta <1\text{),}  \label{42.41}
\end{equation}
where the coefficient $a$ and the exponent $\alpha $ are arbitrary positive
parameters. The members of this class are fixed points of \emph{power-law}
Poissonian renormalizations. For this class:

\begin{enumerate}
\item[$\bullet $] The maximal distribution is characterized by the CDF 
\begin{equation}
F_{\text{max}}(\theta )=\exp \left\{ -a\left( -\ln (\theta )\right) ^{\alpha
}\right\}  \label{42.42}
\end{equation}
($0<\theta <1$).

\item[$\bullet $] The minimal distribution is degenerate: the minimum $X_{%
\text{min}}$ equals zero with probability one.

\item[$\bullet $] The aggregate distribution is the CS-fractal distribution,
characterized by the CS 
\begin{equation}
C_{\text{agg}}(n)=\frac{\Gamma (1+\alpha )a}{n^{\alpha }}  \label{42.43}
\end{equation}
($n=1,2,\cdots $).
\end{enumerate}

The CDF of equation (\ref{42.42}) reduces to the \emph{Beta} CDF when
setting the exponent value $\alpha $ to unity. Thus, we refer to the
distribution corresponding to this CDF as \emph{Hyper Beta}.

\subsection{\label{4.3}Structural properties of Poissonian fractals}

In this Subsection we describe the structural properties of the four classes
of Poissonian fractals presented in the previous Subsection.

\subsubsection{Power-law structure and intrinsic scales}

The CRFs of the linear and log-linear Poissonian fractals admit the
power-law structure 
\begin{equation}
R(\cdot )=a\left( \overset{\text{ }}{S(\cdot )}\right) ^{\alpha }\text{ ,}
\label{43.11}
\end{equation}
where the coefficient $a$ and the exponent $\alpha $ are arbitrary positive
parameters, and where the function $S(\cdot )$ is the \emph{intrinsic scale}
of the class under consideration:

\begin{enumerate}
\item[$\bullet $] \emph{Linear scale} $S(\theta )=\theta $ in the case of
linear Poissonian fractals ($\theta >0$).

\item[$\bullet $] \emph{Log-linear scale} $S(\theta )=\ln (\theta )$ in the
case of log-linear Poissonian fractals ($\theta >1$).
\end{enumerate}

Analogously, the SRFs of harmonic and log-harmonic Poissonian fractals admit
the power-law structure 
\begin{equation}
\overline{R}(\cdot )=a\left( \overset{\text{ }}{S(\cdot )}\right) ^{\alpha }%
\text{ ,}  \label{43.12}
\end{equation}
where the coefficient $a$ and the exponent $\alpha $ are arbitrary positive
parameters, and where the function $S(\cdot )$ is the \emph{intrinsic scale}
of the class under consideration:

\begin{enumerate}
\item[$\bullet $] \emph{Harmonic scale} $S(\theta )=\theta ^{-1}$ in the
case of harmonic Poissonian fractals ($\theta >0$).

\item[$\bullet $] \emph{Log-harmonic scale }$S(\theta )=\ln \left( \theta
^{-1}\right) $ in the case of log-harmonic Poissonian fractals ($0<\theta <1$%
).
\end{enumerate}

\bigskip

All four classes of Poissonian fractals share the \emph{common} power-law
structure $y=ax^{\alpha }$. What distinguishes one fractal class from
another is the \emph{intrinsic scale} -- on which the power-law structure is
composed.

\subsubsection{Order statistics and Exponential representations}

All four classes of Poissonian fractals posses an underlying \emph{%
Exponential structure} which we now describe. Let $\left\{ \mathcal{E}%
_{n}\right\} _{n=1}^{\infty }$ denote an IID sequence of
Exponentially-distributed random variables with unit mean.

The points of linear and log-linear Poissonian fractals can be listed in an
increasing order $\xi _{1}<\xi _{2}<\xi _{3}<\cdots $. The \emph{order
statistics} $\left\{ \xi _{n}\right\} _{n=1}^{\infty }$, in turn, admit the
following exponential representations:

\begin{enumerate}
\item[$\bullet $] Linear Poissonian fractals: 
\begin{equation}
\xi _{n}\overset{\text{Law}}{=}\left( \frac{\mathcal{E}_{1}+\cdots +\mathcal{%
E}_{n}}{a}\right) ^{1/\alpha }  \label{43.21}
\end{equation}%
($n=1,2,\cdots $). Equation (\ref{43.21}) is the infinite-dimensional
counterpart of the one-dimensional exponential representations of the Fr\'{e}%
chet distribution given in equation (\ref{22.31}).

\item[$\bullet $] Log-linear Poissonian fractals: 
\begin{equation}
\xi _{n}\overset{\text{Law}}{=}\exp \left\{ \left( \frac{\mathcal{E}%
_{1}+\cdots +\mathcal{E}_{n}}{a}\right) ^{1/\alpha }\right\}   \label{43.22}
\end{equation}%
($n=1,2,\cdots $). Equation (\ref{43.22}) is the infinite-dimensional
generalization of the one-dimensional exponential representations of the
Pareto distribution given in equation (\ref{21.31}).
\end{enumerate}

Analogously, the points of harmonic and log-harmonic Poissonian fractals can
be listed in a decreasing order $\xi _{1}>\xi _{2}>\xi _{3}>\cdots $. The 
\emph{order statistics} $\left\{ \xi _{n}\right\} _{n=1}^{\infty }$, in
turn, admit the following exponential representations:

\begin{enumerate}
\item[$\bullet $] Harmonic Poissonian fractals: 
\begin{equation}
\xi _{n}\overset{\text{Law}}{=}\left( \frac{\mathcal{E}_{1}+\cdots +\mathcal{%
E}_{n}}{a}\right) ^{-1/\alpha }  \label{43.23}
\end{equation}%
($n=1,2,\cdots $). Equation (\ref{43.23}) is the infinite-dimensional
counterpart of the one-dimensional exponential representations of the
Weibull distribution given in equation (\ref{22.31}).

\item[$\bullet $] Log-harmonic Poissonian fractals: 
\begin{equation}
\xi _{n}\overset{\text{Law}}{=}\exp \left\{ -\left( \frac{\mathcal{E}%
_{1}+\cdots +\mathcal{E}_{n}}{a}\right) ^{1/\alpha }\right\}   \label{43.24}
\end{equation}%
($n=1,2,\cdots $). Equation (\ref{43.24}) is the infinite-dimensional
generalization of the one-dimensional exponential representations of the
Beta distribution given in equation (\ref{21.31}).
\end{enumerate}

Equations (\ref{43.21})-(\ref{43.24}) follow from the \textquotedblleft
displacement theorem\textquotedblright\ of the theory of Poisson processes (%
\cite{Kin}, Section \emph{5.5}), combined with the fact that the increasing
sequence $\left\{ \mathcal{E}_{1}+\cdots +\mathcal{E}_{n}\right\}
_{n=1}^{\infty }$ forms a standard unit-rate Poisson process.

The reciprocal connection between equations (\ref{43.21}) and (\ref{43.23})
is the infinite-dimensional counterpart of the one-dimensional reciprocal
connection between the Fr\'{e}chet and Weibull distributions given by
equation (\ref{22.32}); the reciprocal connection between equations (\ref%
{43.22}) and (\ref{43.24}) is the infinite-dimensional generalization of the
one-dimensional reciprocal connection between the Beta and Pareto
distributions given by equation (\ref{21.32}).

\subsubsection{Transforming between fractal classes}

It is possible to transform from an ``input'' Poissonian fractal $\mathcal{X}
$ belonging to one fractal class to an ``output'' Poissonian fractal $%
\mathcal{Y}$ belonging to another fractal class via a simple point-to-point
mapping $x\mapsto y=\psi (x)$ -- which transforms the points $x$ of the
``input'' Poissonian fractal to the points $y$ of the ``output'' Poissonian
fractal.

The point-to-point mappings are given in Table \emph{2}, which should be
read as follows: in order to transform from the fractal class of row $i$ to
the fractal class of column $j$ one has to apply the point-to-point mapping $%
y=\psi (x)$ appearing in cell $(i,j)$ of the Table. The construction of
these point-to-point mappings follows straightforwardly from the
\textquotedblleft displacement theorem\textquotedblright\ of the theory of
Poisson processes (\cite{Kin}, Section \emph{5.5}).

Each class of Poissonian fractals has one degenerate extremal and one
non-degenerate extremal -- see Table \emph{3 }below. The point-to-point
mappings of Table \emph{2} transform the degenerate extremals amongst
themselves, and transform the non-degenerate extremals amongst themselves.
The point-to-point mappings of Table \emph{2} do \emph{not}, however,
transform the aggregates of the different Poissonian fractal classes to each
other.

\begin{center}
{\large Table 2}

\textbf{Point-to-point mappings of the Poissonian fractal classes}

\bigskip

\begin{tabular}{||c||c||c||c||c||}
\hline\hline
& $%
\begin{array}{c}
\text{\textbf{\ }} \\ 
\text{\textbf{Lin.}} \\ 
\text{\textbf{\ }}%
\end{array}
$ & $%
\begin{array}{c}
\text{\textbf{\ }} \\ 
\text{\textbf{Har.}} \\ 
\text{\textbf{\ }}%
\end{array}
$ & $%
\begin{array}{c}
\text{\textbf{\ }} \\ 
\text{\textbf{Log-lin.}} \\ 
\text{\textbf{\ }}%
\end{array}
$ & $%
\begin{array}{c}
\text{\textbf{\ }} \\ 
\text{\textbf{Log-har.}} \\ 
\text{\textbf{\ }}%
\end{array}
$ \\ \hline\hline
$%
\begin{array}{c}
\text{\textbf{\ }} \\ 
\text{\textbf{Lin.}} \\ 
\text{\textbf{\ }}%
\end{array}
$ & $%
\begin{array}{c}
\text{ } \\ 
x \\ 
\text{ }%
\end{array}
$ & $%
\begin{array}{c}
\text{ } \\ 
x^{-1} \\ 
\text{ }%
\end{array}
$ & $%
\begin{array}{c}
\text{ } \\ 
\exp (x) \\ 
\text{ }%
\end{array}
$ & $%
\begin{array}{c}
\text{ } \\ 
\exp (-x) \\ 
\text{ }%
\end{array}
$ \\ \hline\hline
$%
\begin{array}{c}
\text{\textbf{\ }} \\ 
\text{\textbf{Har.}} \\ 
\text{\textbf{\ }}%
\end{array}
$ & $%
\begin{array}{c}
\text{ } \\ 
x^{-1} \\ 
\text{ }%
\end{array}
$ & $%
\begin{array}{c}
\text{ } \\ 
x \\ 
\text{ }%
\end{array}
$ & $%
\begin{array}{c}
\text{ } \\ 
\exp (x^{-1}) \\ 
\text{ }%
\end{array}
$ & $%
\begin{array}{c}
\text{ } \\ 
\exp (-x^{-1}) \\ 
\text{ }%
\end{array}
$ \\ \hline\hline
$%
\begin{array}{c}
\text{\textbf{\ }} \\ 
\text{\textbf{Log-lin.}} \\ 
\text{\textbf{\ }}%
\end{array}
$ & $%
\begin{array}{c}
\text{ } \\ 
\ln (x) \\ 
\text{ }%
\end{array}
$ & $%
\begin{array}{c}
\text{ } \\ 
\left( \ln (x)\right) ^{-1} \\ 
\text{ }%
\end{array}
$ & $%
\begin{array}{c}
\text{ } \\ 
x \\ 
\text{ }%
\end{array}
$ & $%
\begin{array}{c}
\text{ } \\ 
x^{-1} \\ 
\text{ }%
\end{array}
$ \\ \hline\hline
$%
\begin{array}{c}
\text{\textbf{\ }} \\ 
\text{\textbf{Log-har.}} \\ 
\text{\textbf{\ }}%
\end{array}
$ & $%
\begin{array}{c}
\text{ } \\ 
-\ln (x) \\ 
\text{ }%
\end{array}
$ & $%
\begin{array}{c}
\text{ } \\ 
-\left( \ln (x)\right) ^{-1} \\ 
\text{ }%
\end{array}
$ & $%
\begin{array}{c}
\text{ } \\ 
x^{-1} \\ 
\text{ }%
\end{array}
$ & $%
\begin{array}{c}
\text{ } \\ 
x \\ 
\text{ }%
\end{array}
$ \\ \hline\hline
\end{tabular}
\end{center}

\bigskip

\begin{center}
{\large Table 3}

\textbf{Extremals of the Poissonian fractal classes}

\bigskip

\begin{tabular}{||c||c||c||}
\hline\hline
$%
\begin{array}{c}
\text{\textbf{\ }} \\ 
\text{\textbf{Fractal}} \\ 
\text{\textbf{Class}} \\ 
\text{\textbf{\ }}%
\end{array}%
$ & $%
\begin{array}{c}
\text{\textbf{\ }} \\ 
\text{\textbf{Degenerate}} \\ 
\text{\textbf{extremals}} \\ 
\text{\textbf{\ }}%
\end{array}%
$ & $%
\begin{array}{c}
\text{\textbf{\ }} \\ 
\text{\textbf{Non-degenerate}} \\ 
\text{\textbf{extremals}} \\ 
\text{\textbf{\ }}%
\end{array}%
$ \\ \hline\hline
$%
\begin{array}{c}
\text{\textbf{\ }} \\ 
\text{\textbf{Linear}} \\ 
\text{\textbf{\ }}%
\end{array}%
$ & $%
\begin{array}{c}
\text{ } \\ 
\text{Max}=\infty  \\ 
\text{ }%
\end{array}%
$ & $%
\begin{array}{c}
\text{ } \\ 
\text{Min}\sim \text{\emph{Weibull}} \\ 
\text{ }%
\end{array}%
$ \\ \hline\hline
$%
\begin{array}{c}
\text{\textbf{\ }} \\ 
\text{\textbf{Harmonic}} \\ 
\text{\textbf{\ }}%
\end{array}%
$ & $%
\begin{array}{c}
\text{ } \\ 
\text{Min}=0 \\ 
\text{ }%
\end{array}%
$ & $%
\begin{array}{c}
\text{ } \\ 
\text{Max}\sim \text{\emph{Fr\'{e}chet}} \\ 
\text{ }%
\end{array}%
$ \\ \hline\hline
$%
\begin{array}{c}
\text{\textbf{\ }} \\ 
\text{\textbf{Log-linear}} \\ 
\text{\textbf{\ }}%
\end{array}%
$ & $%
\begin{array}{c}
\text{ } \\ 
\text{Max}=\infty  \\ 
\text{ }%
\end{array}%
$ & $%
\begin{array}{c}
\text{ } \\ 
\text{Min}\sim \text{\emph{Hyper Pareto}} \\ 
\text{ }%
\end{array}%
$ \\ \hline\hline
$%
\begin{array}{c}
\text{\textbf{\ }} \\ 
\text{\textbf{Log-harmonic}} \\ 
\text{\textbf{\ }}%
\end{array}%
$ & $%
\begin{array}{c}
\text{ } \\ 
\text{Min}=0 \\ 
\text{ }%
\end{array}%
$ & $%
\begin{array}{c}
\text{ } \\ 
\text{Max}\sim \text{\emph{Hyper Beta}} \\ 
\text{ }%
\end{array}%
$ \\ \hline\hline
\end{tabular}
\end{center}

\newpage 

\section{Conclusions}

In this paper we examined the definition of fractality in the context of
positive-valued probability distributions. We followed the conventional
approach of associating the notion of fractality with power-law structures
-- considering various distribution characteristics, rather than the
survival probability alone.

We proved the existence of no less than six different classes of fractal
probability distributions -- all admitting a characteristic power-law
structure, and all being the unique fixed-points of renormalizations acting
on positive-valued probability distributions. Each class manifested a
markedly different meaning of fractality.

All fractal classes were further shown to admit an underlying Poissonian
structure -- each fractal distribution being a one-dimensional functional
projection of an underlying Poisson process. The underlying Poisson
processes, in turn, are fractal objects -- being the unique fixed-points of
Poissonian renormalizations.

The notion of fractality on the one-dimensional ``probability-distribution
level'' emanated from the notion of fractality on the infinite-dimensional
``Poisson-process level'':

On the ``probability-distribution level'' fractality was defined
algebraically via power-law structures, and the connection between the
different classes of fractal distributions was unclear.

On the elemental ``Poisson-process level'', however, fractality was defined
via the geometric notion of population-renormalization, and a unified
picture of fractality was obtained: it became vividly clear how all classes
of fractal distributions emerge from the underlying Poissonian fractals, how
they connect to each other, and how the underlying Poissonian fractals
connect to each other.

We have seen that on the elemental ``Poisson-process level'', fractals do
admit a universal power-law structure, yet this structure is intertwined
with another key structure: the intrinsic scale which can be either linear,
harmonic, log-linear, or log-harmonic. Whereas the power-law structure is
common to all Poissonian fractals, it is the intrinsic scale which
differentiates between the Poissonian fractal classes and characterizes them.

This research provides a panoramic and comprehensive view of fractal
distributions, backed by a unified theory of their underlying Poissonian
fractals.

\newpage

\section{Appendix: proof of equation (\protect\ref{23.42})}

We compute the LLT $\Psi _{D}(\theta )$ ($\theta \geq 0$) corresponding to
the power-law CS $C_{D}(n)=an^{-\alpha }$ ($n=1,2,\cdots $).

\bigskip

Note that 
\begin{equation}
C_{D}(n)=\frac{a}{n^{\alpha }}=a\int_{0}^{\infty }\exp \left\{ -nt\right\} 
\frac{t^{\alpha -1}}{\Gamma (\alpha )}dt  \label{A11}
\end{equation}
($n=1,2,\cdots $). Hence, substituting equation (\ref{A11}) into equation (%
\ref{23.41}) gives 
\begin{equation}
\left. 
\begin{array}{l}
\Psi _{D}(\theta )=-\sum_{n=1}^{\infty }C_{D}(n)\frac{(-\theta )^{n}}{n!} \\ 
\\ 
=-\sum_{n=1}^{\infty }\left( a\int_{0}^{\infty }\exp \left\{ -nt\right\} 
\frac{t^{\alpha -1}}{\Gamma (\alpha )}dt\right) \frac{(-\theta )^{n}}{n!} \\ 
\\ 
=-\int_{0}^{\infty }\left( \sum_{n=1}^{\infty }\frac{(-\theta \exp \left\{
-t\right\} )^{n}}{n!}\right) \left( \frac{a}{\Gamma (\alpha )}t^{\alpha
-1}\right) dt \\ 
\\ 
=\int_{0}^{\infty }\left( \overset{\text{ }}{1-\exp \left\{ -\theta \exp
\left\{ -t\right\} \right\} }\right) \left( \frac{a}{\Gamma (\alpha )}%
t^{\alpha -1}\right) dt%
\end{array}
\right.  \label{A12}
\end{equation}
($\theta \geq 0$).

Now: 
\begin{equation}
\Psi _{D}(\theta )=\int_{0}^{\infty }\left( \overset{\text{ }}{1-\exp
\left\{ -\theta \exp \left\{ -t\right\} \right\} }\right) \left( \frac{a}{%
\Gamma (\alpha )}t^{\alpha -1}\right) dt  \label{A13}
\end{equation}
(using the change of variables $u=\exp \left\{ -t\right\} $) 
\begin{equation}
=\int_{0}^{1}\left( \overset{\text{ }}{1-\exp \left\{ -\theta u\right\} }%
\right) \left( \frac{a}{\Gamma (\alpha )}\frac{\left( -\ln (u)\right)
^{\alpha -1}}{u}\right) du  \label{A14}
\end{equation}
(using integration by parts) 
\begin{equation}
=\int_{0}^{1}\theta \exp \left\{ -\theta x\right\} \left( \int_{x}^{1}\frac{a%
}{\Gamma (\alpha )}\frac{\left( -\ln (u)\right) ^{\alpha -1}}{u}du\right) dx%
\text{ .}  \label{A15}
\end{equation}

On the other hand (using the change of variables $t=-\ln (u)$) we have 
\begin{equation}
\int_{x}^{1}\frac{a}{\Gamma (\alpha )}\frac{\left( -\ln (u)\right) ^{\alpha
-1}}{u}du=\int_{0}^{-\ln (x)}\frac{a}{\Gamma (\alpha )}t^{\alpha -1}dt=\frac{%
a}{\Gamma (1+\alpha )}\left( -\ln (x)\right) ^{\alpha }\text{ .}  \label{A16}
\end{equation}
Hence, substituting equation (\ref{A16}) into equation (\ref{A15}) we
conclude that 
\begin{equation}
\Psi _{D}(\theta )=\theta \int_{0}^{1}\exp \left\{ -\theta x\right\} \left( 
\frac{a}{\Gamma (1+\alpha )}\left( -\ln (x)\right) ^{\alpha }\right) dx
\label{A17}
\end{equation}
($\theta \geq 0$).

\end{document}